\documentclass[%
reprint,
superscriptaddress,
nobibnotes,
amsmath,amssymb,
aps,
prl,
]{revtex4-2}

\usepackage{graphicx}% Include figure files
\usepackage{dcolumn}% Align table columns on decimal point
\usepackage{bm}% bold math
\usepackage{hyperref}% add hypertext capabilities
\usepackage{siunitx}
\usepackage{here}
\usepackage{url}
\usepackage[mathlines]{lineno}% Enable numbering of text and display math

\newcommand{\KEK}{High Energy Accelerator Research Organization, 1-1 Oho, Tsukuba, Ibaraki 305-0801, Japan}
\newcommand{\IbarakiUniv}{Graduate School of Science and Engineering, Ibaraki University, Mito, Ibaraki 310-8512, Japan}
\newcommand{\UnivofTokyo}{Graduate School of Science, University of Tokyo, 7-3-1 Hongo, Bunkyo-ku, Tokyo 113-0033, Japan}
\newcommand{\SOKENDAI}{The Graduate University for Advanced Studies, Kanagawa 240-0193, Japan}
\newcommand{\CSNS}{China Spallation Neutron Source Science Center, Dongguan 523803, China}
\newcommand{\IHEP}{Institute of High Energy Physics, Chinese Academy of Sciences, Beijing 100049, China}

\begin{document}

\preprint{APS/123-QED}

\title{
    First Experimental Demonstration of Beam Storage \\ 
    by Three-Dimensional Spiral Injection Scheme
    for Ultra-Compact Storage Rings
}% Force line breaks with \\

\author{R.~Matsushita}\email{Contact author: matsur@post.kek.jp}\affiliation{\UnivofTokyo}\affiliation{\KEK}
\author{H.~Iinuma}\email{Contact author: hiromi.iinuma.spin@vc.ibaraki.ac.jp}\affiliation{\IbarakiUniv}
\author{S.~Ohsawa}\affiliation{\KEK}
\author{H.~Nakayama}\affiliation{\KEK}
\author{K.~Furukawa}\affiliation{\KEK}
\author{S.~Ogawa}\affiliation{\KEK}
\author{N.~Saito}\affiliation{\UnivofTokyo}\affiliation{\KEK}
\author{T.~Mibe}\affiliation{\UnivofTokyo}\affiliation{\KEK}
\author{M.~A.~Rehman}\thanks{Present address: \CSNS, \IHEP}\affiliation{\SOKENDAI}

% \date{\today}% It is always \today, today,
%  but any date may be explicitly specified

\begin{abstract}
Three-dimensional spiral injection enables beam storage in ultra-compact rings with nanosecond revolution periods. 
We report first storage of a $297 \,\si{keV/}c$ electron beam in a $22 \,\si{cm}$ weak-focusing ring 
with a $4.7\,\si{ns}$ revolution period using a $140\,\si{ns}$ kicker pulse. 
A scintillating-fiber detector observes signals $>5\sigma$ above noise for $\geq 1\,\si{\mu s}$, 
and varying the weak-focusing field potential shifts the stored-beam region, consistent with Monte Carlo predictions, validating beam storage.
This proof-of-principle opens a path to ultra-compact storage rings for next-generation precision measurements.
\end{abstract}

\maketitle

% 
% Introduction
% 
% \linenumbers
%
{\it Introduction---}
Storage rings have played a central role in accelerator science 
since their first realization as colliders in the 1960s~\cite{AdA, VEP-1, Princeton}. 
Their applications now span a wide range of fields, including particle and nuclear physics,
synchrotron light sources, medical treatments, and various industrial applications. 
As these applications have expanded, injection methods have undergone continuous development. 
Conventional transverse injection relies on a pulsed kicker magnet 
that produces a localized dipole field which steers the injected beam toward the stored orbit. 
This traditional two-dimensional approach has matured over several decades 
and evolved into the well-established local bump orbit technique~\cite{Lee_Textbook}. 
Ongoing research has also generated next-generation concepts 
such as nonlinear or multipole kicker injection~\cite{PhysRevSTAB.10.123501, PhysRevSTAB.13.020705}, 
swap-out injection~\cite{Swap-out}, and longitudinal injection~\cite{PhysRevSTAB.18.020701}. 
These developments have largely been motivated by advanced light-source facilities 
where ultralow emittance of storage beam requires extremely precise control 
and injection into very narrow dynamic apertures~\cite{Aiba_Review}.

The present study addresses a different direction of progress. 
We focus on the ultra-compact storage rings, since a smaller ring reduces the region 
in which the experimental environment must be controlled for physics measurements,
for example with respect to magnetic field uniformity and the installation of large-scale detector components.
These features are especially important for next-generation precision-measurement experiments. 

\begin{figure*}[tbp]
    \centering
    \includegraphics[width=\linewidth]{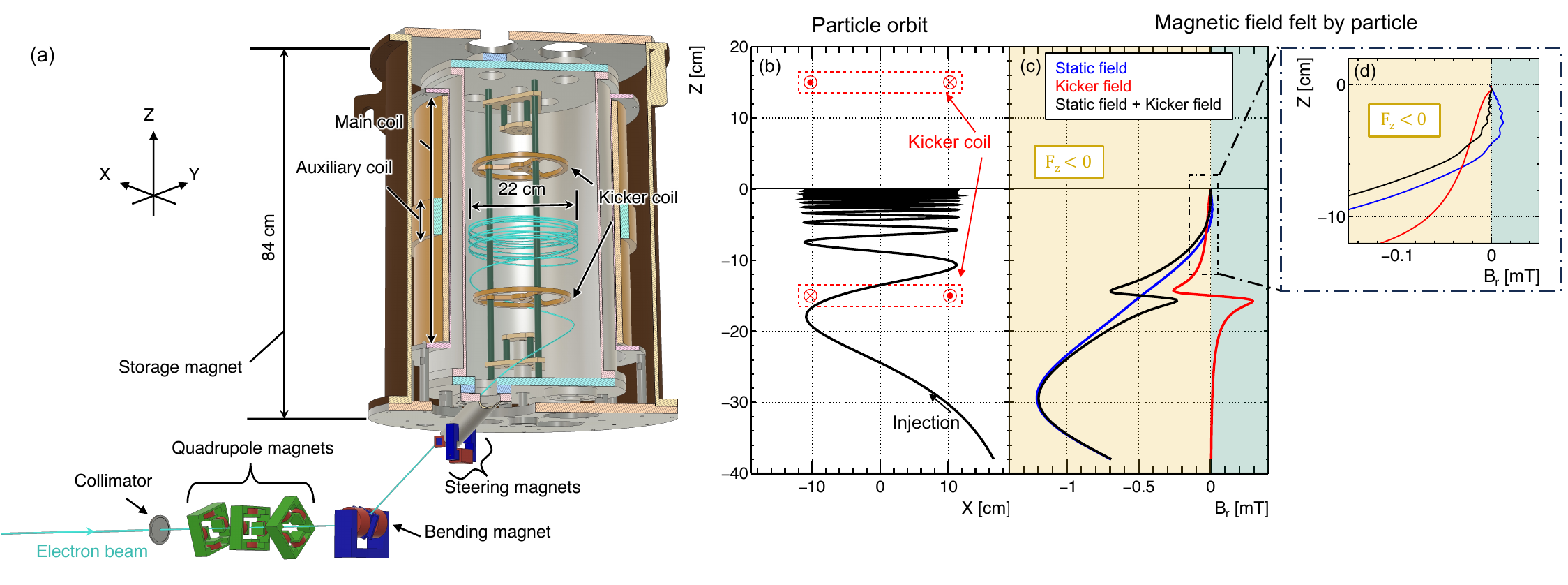}
    \caption{
        (a): Overview of the demonstration beamline. 
        The electron beam is generated by an electron gun and injected into the storage magnet 
        through a transport line consisting of three rotatable quadrupole magnets, 
        one bending magnet, 
        and a pair of steering magnets.
        (b) X-Z projection of the reference particle orbit after injection. 
        The storage midplane corresponds to $Z = 0\,\si{cm}$.
        (c),(d) $B_r$ components along the reference particle orbit. 
        Blue line shows the static magnetic field by the storage magnet. 
        Red line shows the time-varying magnetic field by the kicker.
        The $B_r < 0$ field during injection provides the downward steering to electrons
        required for beam storage in weak-focusing region.
        }
    \label{fig:SITE_overview}
\end{figure*}

However, ultra-compact storage rings face a fundamental difficulty. 
As the size of the ring becomes smaller, the revolution period of the circulating beam decreases to a few nanoseconds. 
Under these conditions, conventional injection schemes become challenging 
because they require strong pulsed magnetic fields with nanosecond-scale rise and fall times. 
Even state-of-the-art developments for the ILC damping ring demonstrated switching times of approximately $3\,\si{ns}$~\cite{ILC_Kicker}
, and achieving significantly faster performance remains technically demanding. 
This limitation has formed a major obstacle in the realization of ultra-compact storage rings.
For nanosecond revolution period, injection with minimal perturbation requires a kick confined to single turn, 
placing simultaneous demands on both switching speed and peak field that are difficult to meet with conventional pulsed-power systems.

A concept proposed in 2016 by {\it H.~Iinuma et al.}~\cite{H.Iinuma2016} offers a solution.
The idea is known as the {\it three-dimensional spiral injection scheme}, 
and it avoids the time-scale limitation by extending the injection orbit into the vertical direction. 
Instead of injecting the beam parallel to the solenoidal axis, the beam enters the solenoid-type storage magnet with a controlled pitch angle. 
This non-parallel injection causes the beam to experience the fringe field of the solenoid, which provides an initial vertical steering force. 
The storage region is defined by a weak-focusing potential~\cite{WF_RRWilson, WF_DWKerst} located near the mid-plane of the magnet. 
As the injected beam enters the weak-focusing potential region, the radial field component $B_r$ changes sign due to the weak-focusing field, which reverses the vertical Lorentz force and steers the beam upward.
To counteract this effect, the spiral injection concept employs a pulsed kicker and this supplies a small vertical kick to reduces pitch angle during many consecutive turns. 

The accumulated effect of these repeated vertical kicks guides the injected beam into the weak-focusing potential, 
allowing stable storage. Because the required deflection is distributed over multiple turns, 
the method avoids the need for the extremely fast pulsed-power performance that limits conventional two-dimensional injection.

The ability to implement this scheme paves the way toward ultra-compact storage rings that are impractical with conventional injection schemes.
The ultra-compact rings are essential for precision measurements involving short-lived particles.
The muon $g-2$/EDM experiment at J-PARC~\cite{E34} and the muEDM experiment at PSI~\cite{muEDM} require the storage of relativistic beams in regions only a few tens of centimeters across, with revolution periods of only a few nanoseconds. 
Compact rings enable relativistic beams to be stored in a small, well-controlled volume, 
which is advantageous for precision measurements of short-lived particles.
Beyond muon physics, ultra-compact weak-focusing rings could support high-precision mass spectrometry 
and electric-dipole-moment searches using various particle species.

No experimental demonstration had been reported so far. 
In this Letter we present the first successful storage of an electron beam using this method.

% 
% Experimental setup
% 
{\it Experimental setup}---
A beamline was constructed~\cite{RehmanThesis}
for the demonstration of the three-dimensional spiral injection scheme.
Fig.~\ref{fig:SITE_overview}(a) shows an overview of the demonstration beamline.
An electron beam with momentum is $p=297\,\si{keV/}c$, kinetic energy is $E_{\rm kin}=80\,\si{keV}$ 
and pulsed width is $100\,\si{ns}$ is generated
using a thermionic electron gun~\cite{E-gun} and a chopper system~\cite{R.Matsushita2021}.
The transport line allows us to control the beam phase space~\cite{H.Iinuma2026} and the injection orbit at the injection point. 
This transport line consists of three quadrupole magnets, one bending magnet, 
and a pair of steering magnets~\cite{RehmanThesis}
, then injected into the storage magnet with a $0.7\,\si{rad}$ pitch angle, 
defined with respect to the horizontal (X-Y) plane.

The storage magnet consists of two solenoidal coils~\cite{Rehman_LINAC18}, 
main coil and auxiliary coil shown in Fig.~\ref{fig:SITE_overview}(a),
allow us to apply both solenoidal field and weak-focusing field in the beam storage region.
By setting the coil currents $I_{\rm MAIN}$ and $I_{\rm AUX}$ with opposite polarities,
we adjust the weak-focusing field potential (field index $n\sim 10^{-2}$ in the storage region)
while keeping the solenoidal field constant at $B_z=8.77\,\si{mT}$ on the storage plane ($Z=0\,\si{cm}$).
Fig.~\ref{fig:SITE_overview}(b) shows the projected reference particle orbit in the X-Z plane
and Fig.~\ref{fig:SITE_overview}(c) and (d) show the corresponding $B_r$ component along the reference orbit.
After injection, the beam follows a spiral orbit thanks to the $B_z$ component and simultaneously
being steered downward ($Z < 0$) thanks to the fringe field of $B_r < 0$. 
Inside the weak-focusing region, the $B_r > 0$ from static weak-focusing field applies an upward force. 
To counter this, a $45\,\si{A}$, $140\,\si{ns}$ pulsed current is applied to the kicker coils, 
inductance $L=1.3\,\si{\mu H}$, at $Z = \pm 15\,\si{cm}$, 
generating a $B_r < 0$ field that maintains the downward steering and enables beam storage.

A detector using plastic-scintillating-fiber (SciFi), {\it SciFi-probe}~\cite{R.Matsushita2024}, 
which can be inserted into the storage region, is utilized to measure the storage beams.
The SciFi-probe consists of a $1\,\si{mm}$-diameter and $20\,\si{cm}$-long SciFi,
which is long enough to cover the entire storage region vertically.
The scintillation signal is read out with a photomultiplier tube (PMT).
Because the CSDA range of a $p=297\,\si{keV/}c$ electron in the SciFi is about $100\,\si{\mu m}$
\footnote{The electron CSDA range is obtained from the NIST ESTAR database~\cite{NIST_STAR},
assuming polystyrene (PS) as a material for the scintillating fiber.}, the measurement is destructive.

\begin{figure}[t]
    \centering
    \includegraphics[width=\linewidth]{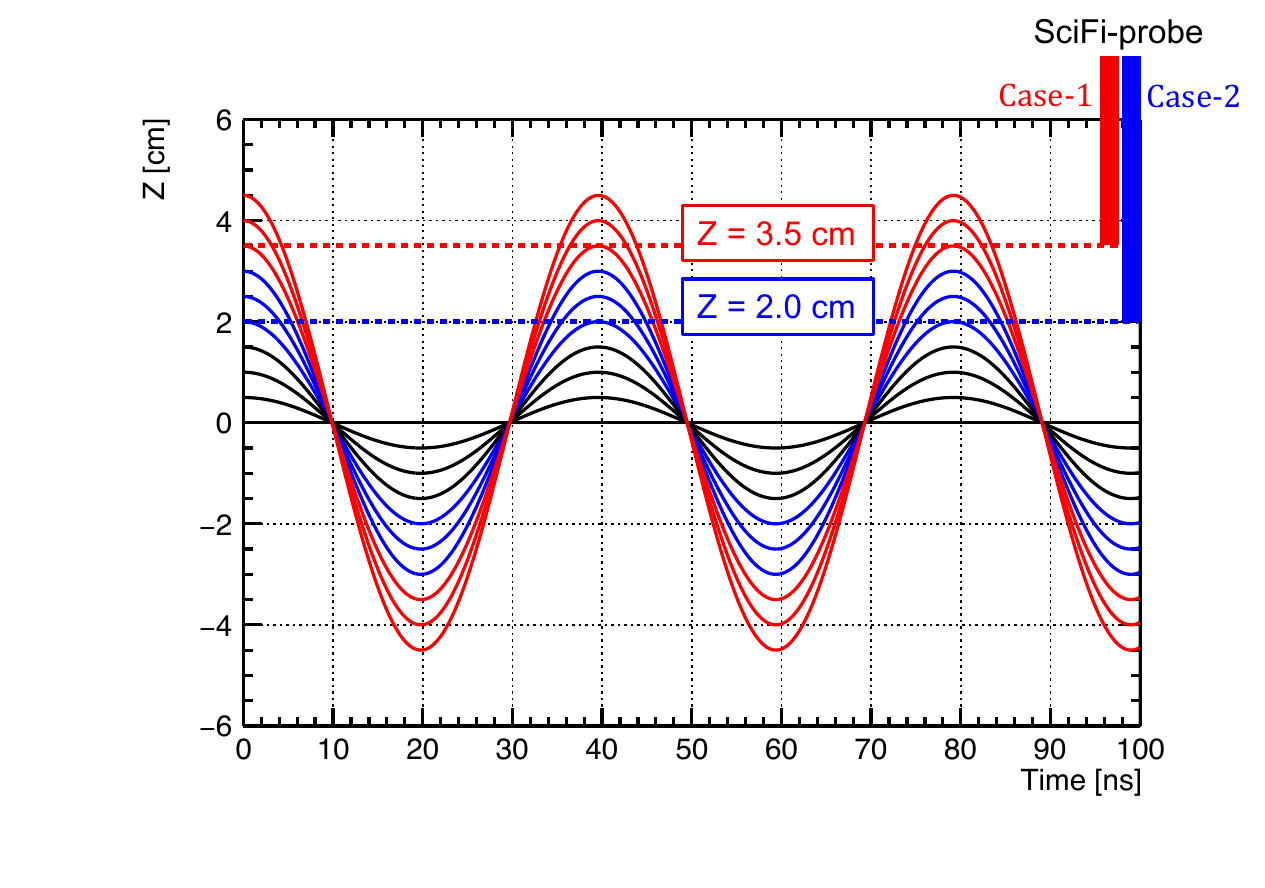}
    \caption{
        Principle of SciFi-probe measurement. 
        Solid lines show the stored particle trajectories with vertical betatron oscillation.
        Stored particles with vertical betatron amplitude greater than $Z\,\si{[cm]}$, 
        which is the tip position of the SciFi-probe, generate signals. 
        Case-1, SciFi-probe is inserted up to $Z = 3.5\,\si{cm}$, measures particles 
        with vertical amplitude $\geqq 3.5\,\si{cm}$ shown by red lines.
        Similarly, Case-2, SciFi-probe is inserted up to $Z = 2.0\,\si{cm}$, 
        measures those with amplitude $\geqq 2.0\,\si{cm}$ shown by red and blue lines.
    }
    \label{fig:SciFiprobe_Meas}
\end{figure}

In a measurement, the SciFi-probe is inserted vertically from above 
with its tip positioned at an insertion depth $Z$, which is measured from the storage midplane ($Z=0\,\si{cm}$),
and particle hits generate a signal.
Because stored particles perform vertical betatron oscillations,
the integrated signal at a given $Z$ is proportional to the population whose vertical amplitude exceeds that depth (Fig.~\ref{fig:SciFiprobe_Meas}).
Since the stored particles also perform betatron oscillations in the radial direction 
simultaneously with vertical betatron oscillations, 
the SciFi-probe does not measure all stored particles within a single vertical betatron oscillation period.
We thus integrate the SciFi-probe signal over a sufficiently long time window to cover a wide range of betatron phases.
In this Letter, we define beam storage operationally as observing a SciFi-probe signal 
that exceeds $5\sigma_{\rm noise}$ for $1\,\si{\micro s}$ or longer, 
which is ten times longer than the injected pulse width of $100\,\si{ns}$ and corresponds to more than $200$ revolutions,
where $\sigma_{\rm noise}$ is the standard deviation (rms fluctuation) of the voltage in a pre-injection time window.
This conservative criterion is introduced to distinguish true storage from non-stored or 
transiently circulating particles that may survive for only a few turns 
before escaping the weak-focusing region.

A plastic scintillator is installed on the bottom surface of the storage chamber 
as a beam loss monitor to observe time structure of reflected beams
that are not captured in the storage region during injection.
Information from this beam loss monitor is used for tuning beam injection trajectory.

\begin{figure}[thb]
    \centering
    \includegraphics[width=\linewidth]{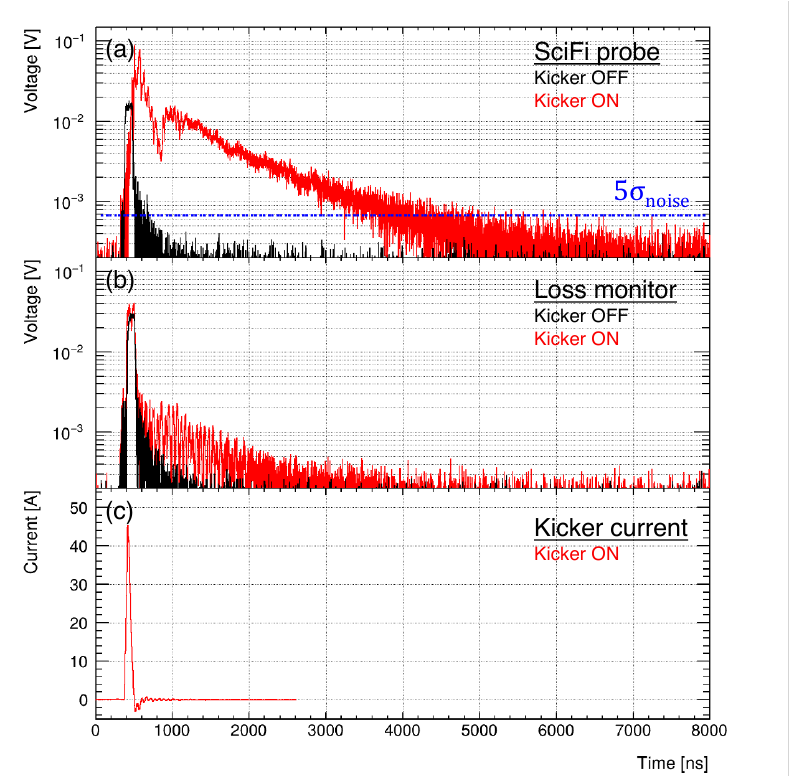}
    \caption{
        Example of measured signals. 
        (a): SciFi-probe signal measured within the weak-focusing region.
        Blue line shows the threshold of $5\sigma_{\rm noise}$.
        (b): Loss-monitor signal downstream of the storage region. 
        (c): Kicker current waveform measured using a Rogowski coil. 
        Red and black lines correspond to the with and without kicker conditions, respectively.
        In this example, the SciFi-probe is inserted up to the storage midplane
        and SciFi covered the range from $Z=0\,\si{cm}$ to $Z=20\,\si{cm}$.
        Each waveform is an average over $1000$ shots.
    }
    \label{fig:MeasSignal}
\end{figure}

\begin{figure*}[tbp]
    \centering
    \includegraphics[width=\linewidth]{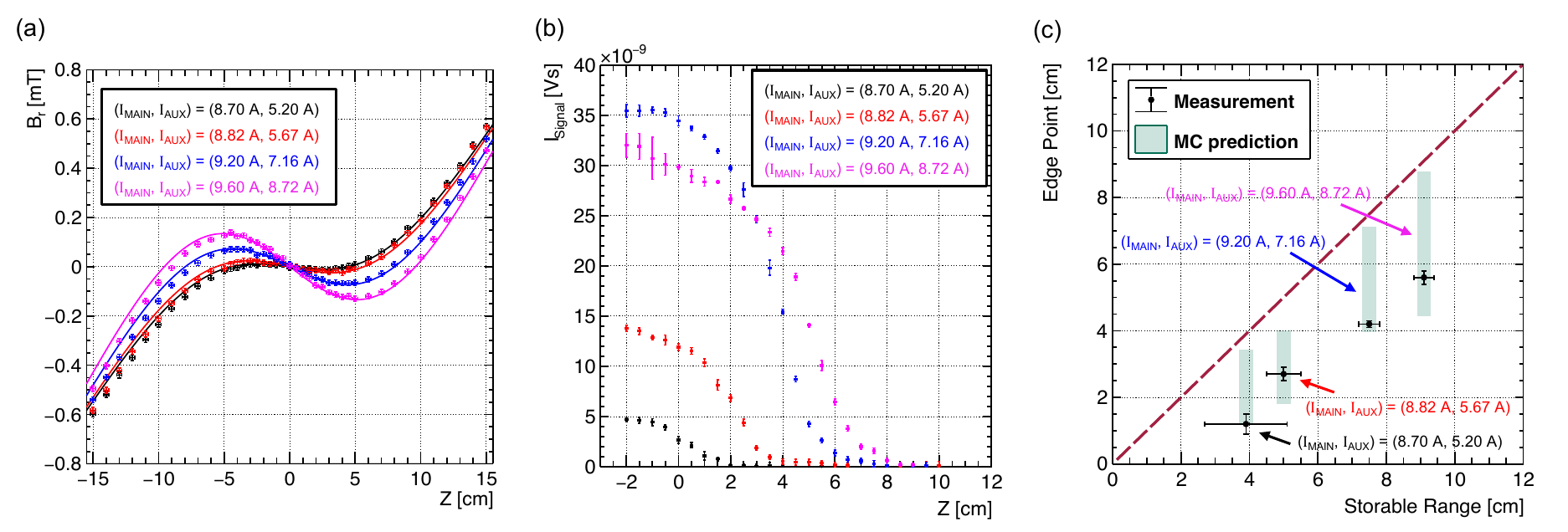}
    \caption{
        Weak-focusing field dependence of the stored beam distribution.
        (a): Weak-focusing field configurations. 
        Colored points denote measurement result, and colored lines denote 
        finite element calculations performed with Opera~\cite{Opera}.
        (b): $Z$-scan results under the four field configurations.
        The vertical axis shows the time-integrated SciFi-probe signal,
        $I_{\rm Signal}=\int V_{\rm Signal}(t)\,dt$,
        where $V_{\rm Signal}(t)$ is the measured voltage as a function of time.
        The integral is evaluated over the time window from $t=560\,\si{ns}$ to $9000\,\si{ns}$,
        starting just after the end of the kicker pulse.
        (c): Comparison of measured stored beam ranges with Monte Carlo predictions. 
        Blue shaded regions show the expected region of MC simulation.
        Black points show the measurement result from field measurement and Z-scan measurement.
        Here the edge point is defined as the $Z$ position where $I_{\rm Signal}$ decreases to $5\%$ of its maximum value.
        For the measured points, the vertical uncertainty is dominated by the SciFi-probe positioning uncertainty and shot-to-shot fluctuations,
        while the horizontal uncertainty is dominated by the magnetic-field measurement accuracy.
    }
    \label{fig:WeakFocusDependency}
\end{figure*}

% 
% Results and Discussions
% 
{\it Experimental results}---
    Beam storage was confirmed using the operational criterion defined above based on the SciFi-probe waveform.
Fig.~\ref{fig:MeasSignal} shows the waveforms from the SciFi-probe, the beam-loss monitor and kicker current for cases with and without the kicker.
With the kicker activated, the SciFi-probe signal remains above $5\sigma_{\rm noise}$ for longer than $1\,\si{\micro s}$, 
thereby satisfying our definition of beam storage and indicating successful beam storage in the weak-focusing region.
The beam-loss monitor simultaneously detected a signal induced by reflected beams, at the expected timing relative to the kicker pulse.
In contrast, without the kicker the SciFi-probe signal has a duration of order $100\,\si{ns}$ comparable to the injected pulse width 
and does not satisfy the $5\sigma_{\rm noise}$ for $1\,\si{\micro s}$ storage criterion, 
indicating that the beam is not stored under these conditions.
This clear contrast demonstrates that three-dimensional spiral injection enables beam storage in ultra-compact storage rings.

During the first several hundred nanoseconds after injection, the stored particles perform coherent vertical betatron oscillations. 
This coherence produced a periodic modulation in the SciFi waveform. Although the modulation originates from the vertical betatron motion, 
its period does not correspond to the betatron oscillation period itself. 
The difference arises because the injected beam has a pulse width that is longer than one betatron period, 
so the SciFi-probe samples particles with a wide distribution of betatron phases within each injection pulse. 
As time evolves, the periodic structure gradually decreases in visibility. 
This reduction is caused by the position dependence of the weak-focusing magnetic field, 
which introduces small variations in the betatron frequency and leads to phase mixing among the stored particles.

The measured storage signal persisted for $\geq 1\,\si{\micro\second}$, as shown in Fig.~\ref{fig:MeasSignal}(a).
This duration is more than one order of magnitude longer than the loss time observed without the kicker and demonstrates 
that multi-turn vertical steering enables stable beam storage even when the revolution period is as short as $4.7\,\si{ns}$.
The observed signal duration is largely limited by scattering in the SciFi-probe, which is inserted into the storage region for measurement.
It therefore provides a lower bound on the intrinsic storage lifetime of the beam, which is expected to be longer than the observed value.

We verified that the accumulated beam is stored in the weak-focusing field by measuring its vertical distribution 
under four different field configurations.
The corresponding magnetic-field distributions are shown in Fig.~\ref{fig:WeakFocusDependency}(a).
For each configuration, we performed a scan of the SciFi-probe insertion depth $Z$, which is measured from the storage midplane, 
hereafter referred to as a {\it $Z$-scan}:
the SciFi-probe was inserted from above in steps of $5\,\si{mm}$, and at each $Z$ we measured a waveform averaged over $1000 \,\si{shots}$.
We then evaluated the time-integrated value $I_{\rm Signal}$, the definition given in the caption of Fig.~\ref{fig:WeakFocusDependency},
which yields the $Z$-scan curve shown in Fig.~\ref{fig:WeakFocusDependency}(b).
In Fig.~\ref{fig:WeakFocusDependency}(c), we define the upper edge of the stored-beam distribution as the $Z$ position 
where $I_{\rm Signal}$ decreases to $5\%$ of its maximum value, 
and compare it with Monte Carlo (MC) predictions of the storable range for each configuration.
The vertical uncertainty of the measured points includes 
(i) the positioning uncertainty of the SciFi-probe 
and (ii) shot-to-shot fluctuations of the $I_{\rm Signal}$,
whereas the horizontal uncertainty is dominated by the measurement accuracy of the weak-focusing field distribution
used to determine the storable range for each configuration.
In the MC predictions, the shaded region represents the spread obtained by including the injection-trajectory uncertainty, 
which is the dominant systematic contribution.
The measurements for all configurations agree with the MC predictions, 
demonstrating that the observed stored-beam distribution is governed by the intended weak-focusing field configuration 
rather than accidental trapping or experimental artifacts.

The present storage efficiency, defined as the fraction of electrons stored from one injected bunch,
is well below $1\%$ under the current injection conditions based on Monte Carlo simulations.
The storage efficiency can be improved by XY-coupled transverse phase-space matching~\cite{H.Iinuma_IPAC25, S.Ogawa_IPAC25}
and a shorter injected bunch length achievable through upstream RF beam manipulation.

% 
% Conclusions
% 
{\it Conclusions}---
We have achieved the first experimental demonstration of beam storage by the three-dimensional spiral injection scheme. 
This demonstration establishes the technical feasibility of this scheme 
and provides a viable pathway toward ultra-compact storage rings 
required for next-generation precision experiments. 
The technique is a key enabling technique for the planned muon $g-2$/EDM experiment at J-PARC
and the muEDM experiment at PSI, 
and may open a new frontier in precision studies of short-lived particles.

{\it Acknowledgments}---
We thank H.~Hisamatsu, H.~Someya, T.~Suwada, Y.~Okayasu, and Y.~Yano for providing equipment used in this work, 
and K.~Sasaki for help with magnetic field measurements. 
We also thank A.~Tokuchi (Pulsed Power Japan Laboratory Ltd.) 
for manufacturing the pulsed-power supply for the kicker system 
and T.~Ushiku (Next Create Service Ltd.) for manufacturing the magnets. 
We acknowledge general support from the KEK Accelerator Injector LINAC group, 
technical support from the KEK Mechanical Engineering Center, 
and assistance with installation work from Futaba Kogyo Co., Ltd. 
This work was supported by JSPS KAKENHI Grants No.~26287055, No.~19H00673, No.~22K14061, and No.~23KJ0590.

\bibliographystyle{apsrev4-2}
\bibliography{main}% Produces the bibliography via BibTeX.
\end{document}